\documentclass[10pt,conference,letterpaper]{IEEEtran}

\usepackage[para]{footmisc}

\usepackage[utf8]{inputenc}
\usepackage[T1]{fontenc}
\usepackage{amsmath,amssymb}
\usepackage{graphicx}
\usepackage{booktabs}
\usepackage{multirow}
\usepackage{xcolor}
\usepackage{url}
\usepackage{hyperref}
\usepackage{pgfplots}
\usepackage{pgfplotstable}
\pgfplotsset{compat=newest}
\usepgfplotslibrary{groupplots,statistics}
\usepackage{tikz}
\usetikzlibrary{patterns,positioning,arrows.meta,calc,shapes.geometric}
\usepackage{subcaption}
\usepackage{balance}
\usepackage{array}
\usepackage{makecell}
\usepackage{pifont}

\usepackage{listings}

\usepackage{cite}
\usepackage{soul}
\usepackage[skip=3pt]{caption}

\definecolor{cprim}{HTML}{0173B2}
\definecolor{ccomp4}{HTML}{DE8F05}
\definecolor{ccomp2}{HTML}{029E73}
\definecolor{ccomp1}{HTML}{D55E00}
\definecolor{heatlow}{HTML}{F7FCF5}
\definecolor{heathigh}{HTML}{00441B}
\definecolor{fitgreen}{HTML}{6BCF7F}
\definecolor{fityellow}{HTML}{FFD93D}
\definecolor{fitred}{HTML}{FF6B6B}

\hypersetup{colorlinks=true,linkcolor=black,citecolor=black,urlcolor=blue}

\begin{document}

\makeatletter
\def\ps@IEEEtitlepagestyle{%
  \def\@oddfoot{}%
  \def\@evenfoot{}%
  \def\@oddhead{%
    \parbox[b]{\textwidth}{\centering
    Author copy of paper published at \textit{34th International Symposium on the Modeling, Analysis, and Simulation of Computer and Telecommunication System (MASCOTS2026)}}
  }%
  \def\@evenhead{\@oddhead}%
}
\makeatother

\title{MCP-GRANITE Benchmark: GRANularity Interface TEsting for MCP-Based LLM Agents} 

\author{
\IEEEauthorblockN{Demetris Paschalides, Moysis Symeonides, George Pallis, Marios D. Dikaiakos}
\IEEEauthorblockA{Department of Computer Science\\
University of Cyprus\\
Nicosia, Cyprus\\
\{dpasch01, msymeo03, pallis, mdd\}@ucy.ac.cy}
}

\maketitle

\begin{abstract}
As LLM agents increasingly interact with external tools through standardized protocols such as MCP, tool-interface design becomes a critical yet underexplored factor. How functionality is decomposed into tools affects whether an agent can select the right tool and construct valid arguments. 
This choice is especially consequential at the edge, where resource constraints limit which models can run locally and scaling up is often not an option. 
We present \textsc{MCP-GRANITE}, an open-source extensible benchmark framework that treats tool-interface granularity as a controlled variable for MCP-based agents, evaluated under edge and IoT scenarios.
It comprises 81 multi-step scenarios across 9 domains, instantiated at 4 granularity levels from fine-grained primitive tools to a single tool. We evaluate 9 locally deployed models (268M-20.9B parameters) across 8,748 trials using task completion, tool selection F1, argument accuracy, latency, and resource-usage metrics. Results show that a 4-tool interface offers the best trade-off, improving task completion by 16.4\% over fine-grained primitives and 33.6\% over a single monolithic tool, while nearly doubling argument accuracy. 
Model size is only weakly correlated with task completion and strongly with latency, while its association with argument accuracy is less robust, and a 3.2B model at the optimal granularity outperforms a 20.9B model at a mismatched one. 
These findings identify tool-interface granularity as a key design parameter for MCP-based agents.
\end{abstract}


\section{Introduction}
\label{sec:intro}

Large language models~(LLMs) are increasingly used as the reasoning core of agentic systems. These systems combine LLM planning with external tool use, such as invoking APIs, reading sensor data, and interacting with services, to ground decisions in executable contexts~\cite{mialon2023augmented, xi2023rise, yao2023react, schick2024toolformer}. This shift creates the need for standardized interfaces through which agents can discover, invoke, and coordinate external tools reliably. The Model Context Protocol~(MCP)~\cite{anthropic2024mcp} has rapidly emerged as the de-facto standard for connecting LLM agents to such capabilities. MCP servers expose tools as named operations, and LLM-based clients discover them at runtime by loading their schemas into the model's context. However, MCP leaves the interface design unspecified, since the same capability can be exposed as many primitive tools, a few task-level functions, a read/write split, or a single monolithic tool. Tool-interface decomposition is, therefore, an explicit design decision that every MCP server developer must make.

This decision matters in practice because the functional decomposition of agentic tools, including their quantity and level of abstraction, directly affects how the agent selects them and populates their arguments. 
In modern application design, the principle that interfaces shape user success is well established. Research on API-usability has shown that abstraction level, naming, and discoverability significantly affect how developers work with software interfaces~\cite{myers2016improving}. Recent work confirms that the same principle holds for LLM agents, as purpose-built tool interfaces substantially improve their performance~\cite{yang2024sweagent}, and even minor changes to the available toolkit's composition can destabilize function calling~\cite{rabinovich2025robustness}. 
Yet despite this evidence, prevailing tool-use benchmarks mainly evaluate whether models can select and invoke tools correctly~\cite{patil2024bfcl, qin2024tool, zhu2024toolsandbox}. Similarly, MCP-specific benchmarks~\cite{yang2025mcpuniverse, song2025mcpgauge} treat the tool interface as fixed and assess only whether the model can use it successfully. 



As a result, these benchmarks overlook the role of the functional decomposition of tools. They do not systematically vary the \textit{interface decomposition} of the same capability set, such as the number of tools, their abstraction level, and their argument structure, as the primary experimental factor. This omission may be less visible for cloud-based LLMs, which can often achieve strong tool-use performance without careful interface design. However, \textit{when increasing model size is not feasible, the tool interface becomes one of the few design parameters available to developers.} 
This is especially relevant at the edge, where LLM agents are increasingly used in IoT and cyber-physical applications under strict latency, energy, and memory constraints~\cite{lin2025edgellmsurvey}. In these settings, tool calls often mediate interactions with local sensors, actuators, and devices, while fixed hardware typically limits the feasible LLM size. 

In such deployments, interface decomposition becomes a practical design knob for edge-hosted LLM agents. Fine-grained interfaces expose many specialized tools with clearer semantics, but increase the number of choices the model must reason over and the call sequences it must compose. Coarse-grained interfaces simplify tool selection, but require the agent to encode richer operational intent through fewer tools with larger and more complex argument schemas. These trade-offs make tool-interface granularity a consequential factor for edge-hosted LLM agents, with measurable effects on performance, robustness, and energy cost. This motivates a controlled benchmark in which researchers and MCP developers can compare alternative interface decompositions of the same capability set while holding the task semantics and execution setting fixed.

In this work, we study how tool-interface granularity shapes the relationship between model scale and performance in constrained environments, making three contributions: (i)~We introduce \textbf{\textsc{MCP-GRANITE}}, an open-source~\cite{edgetoolbench}, extensible benchmark for controlled comparison of MCP tool-interface decompositions. It comprises 81 scenarios across 9 edge/IoT domains, each instantiated at 4 granularity levels
; (ii)~We conduct a \textbf{systematic empirical study} across 9 locally deployed models (268M to 20.9B parameters), showing that the granularity Level~3 interface (4 tools) consistently outperforms both fine-grained (Level~4, 8-10 tools) and highly consolidated (Levels~1-2, 1-2 tools) alternatives, with over-consolidation causing 28.2\% of L1 runs to complete without invoking any tool; and (iii)~We provide a \textbf{scaling and resource-aware analysis} combining functional metrics with GPU utilization and power monitoring, showing that model size is weakly associated with task completion ($\rho{=}0.285$), strongly associated with latency ($\rho{=}0.946$), while its association with argument accuracy ($\rho{=}0.686$) is less robust. Our results show that a 3.2B model at the optimal granularity level outperforms a 20.9B model at a mismatched one, confirming that interface design can outweigh model scale. 

\vspace*{-.2\baselineskip}
\section{Related Work}
\label{sec:related}
\vspace*{-.2\baselineskip}

As tool-augmented LLMs mature~\cite{schick2024toolformer, yao2023react, patil2023gorilla, qin2024tool}, benchmarks develop along two parallel tracks. The first evaluates function-calling correctness, progressing from small tool sets with hierarchical metrics to the Berkeley Function Calling Leaderboard (BFCL)~\cite{patil2024bfcl}, which ranks over 100 models on parallel and nested invocations, and to ToolBench~\cite{qin2024tool} and ToolSandbox~\cite{zhu2024toolsandbox}, which scale multi-step reasoning to real-world and stateful APIs. The second track evaluates broader agent capabilities in interactive environments spanning OS interaction, web tasks, and software engineering~\cite{liu2024agentbench, zhou2024webarena, jimenez2024swebench}. More recently, MCP-specific benchmarks begin evaluating whether models can reliably discover and invoke tools exposed through MCP servers at runtime. MCP-Universe~\cite{yang2025mcpuniverse} reveals that even frontier models achieve only 43.7\% success across 231 real-world MCP tasks, and MCPGAUGE~\cite{song2025mcpgauge} concludes that MCP augmentation does not uniformly improve performance. ToolPlanner~\cite{wu2024toolplanner} and MTU-Bench~\cite{wang2025mtu} vary what they call granularity, but in both cases this refers to the specificity of user instructions or the complexity of evaluation scenarios, not to how the tools themselves are structured. \textit{These efforts collectively raise evaluation realism, yet they share a common assumption, namely that the tool interface is mostly predefined and only model capability is measured against it.}

Moreover, API-usability research has long shown that abstraction level, naming, and discoverability shape how effectively human developers learn and use software interfaces~\cite{myers2016improving}. 
These concerns become more important as pipelines for automatically generating MCP servers from existing APIs become available~\cite{mastouri2026restmcpempiricalstudy}, since design problems in the original API may be transferred directly to agent-facing tool interfaces. Purpose-built tool interfaces, such as SWE-agent~\cite{yang2024sweagent}, substantially improve LLM-agent performance compared with generic alternatives, while even the addition of semantically related tools can destabilize function calling~\cite{rabinovich2025robustness}. 
Recent studies further show that tool descriptions are a critical design surface, with learned rewriting of descriptions improving agent accuracy on unseen tools~\cite{guo2026learning}, while small description edits can disproportionately change tool selection, revealing fragility in agent behavior~\cite{faghih2025toolpreferences}.
These observations become more important under edge constraints, where memory, latency, and energy budgets limit the use of larger models~\cite{lin2025edgellmsurvey}. As small language models can support function calling at the edge, the tool interface becomes a critical design knob when model scaling is not feasible, while the number of available tools affects agentic performance and energy consumption~\cite{Paramanayakam2025}.
Thus, these results establish that what a server exposes, and how it presents functionality to the agent, is not a background condition but a design variable with measurable effects on agent behavior.
Despite this, \textit{no existing benchmark systematically varies the granularity of the same MCP-exposed capability set as the primary experimental factor, particularly for locally deployable models under realistic resource constraints.}

\begin{figure}[t]
    \centering
    \includegraphics[width=0.42\textwidth]{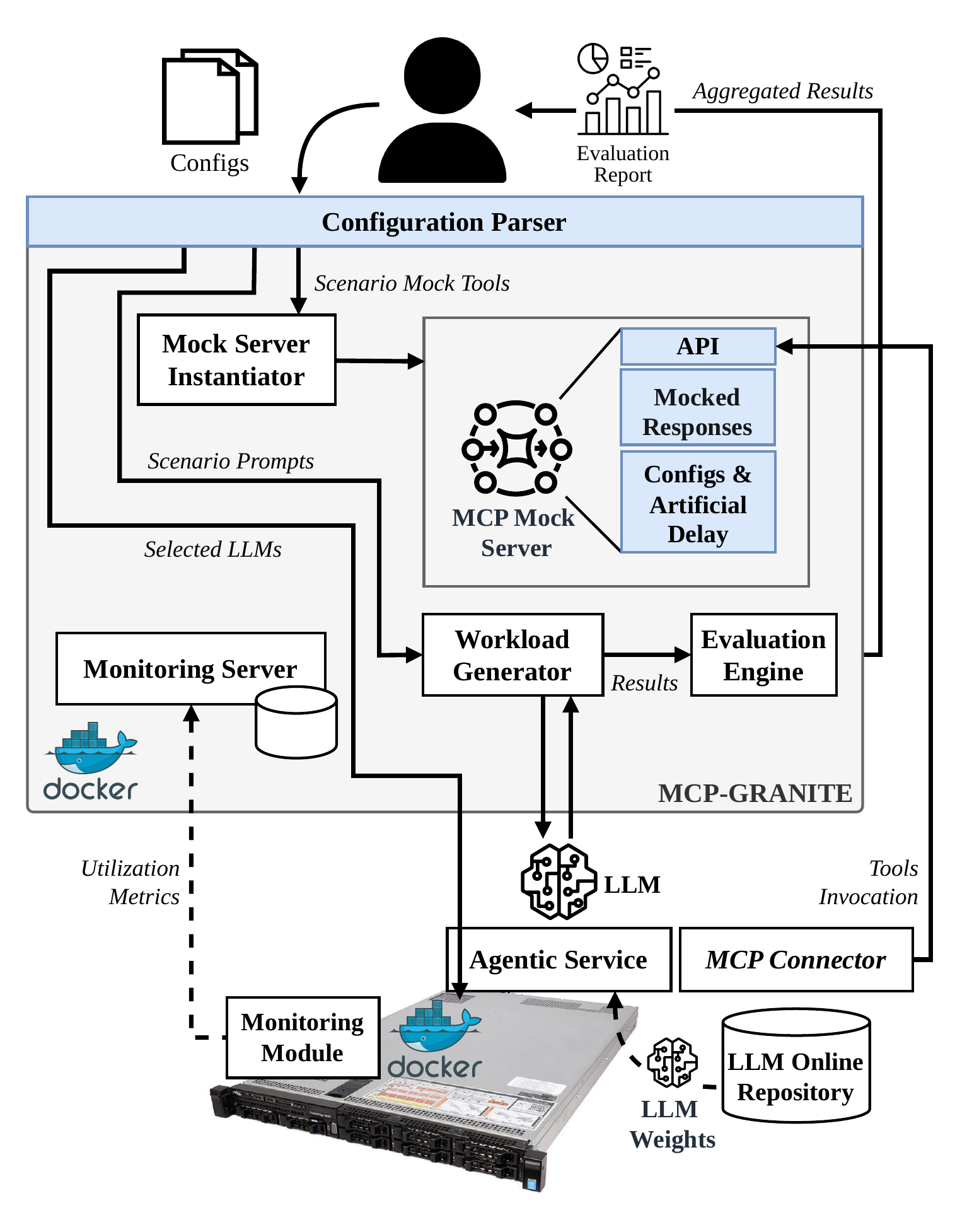}
    \vspace*{-.8\baselineskip}
    \caption{MCP-GRANITE Architecture 
    }
    \label{fig:architecture}
    \vspace*{-1.5\baselineskip}
\end{figure}

\section{The MCP-GRANITE Framework}
\label{sec:design}

\textsc{MCP-GRANITE} addresses the previously-mentioned gap by treating tool-interface decomposition as a core design variable. It is implemented as a controlled execution and evaluation pipeline (Fig.~\ref{fig:architecture}) that separates tool-interface decomposition and experiment specification, tool simulation, agent execution, monitoring, and result aggregation. 
A YAML configuration is resolved by the \emph{Configuration Parser} into the target model, scenario prompts, and the domain- and granularity-specific tool definitions that determine the interface exposed to the agent.
These definitions are forwarded to the \emph{Mock Server Instantiator}, which launches one containerized \emph{MCP Mock Server} per granularity level, each exposing distinct tool schemas over the same domain logic and returning fixed responses, enabling controlled and repeatable evaluation.

On the target edge device, the \emph{Agentic Service} runs in a containerized environment that provides a portable and architecture-agnostic execution substrate. Based on the parsed configuration, the service retrieves the specified LLM weights from HuggingFace and loads the model as the reasoning core of the agent. An MCP connector routes tool invocations from the agent to the corresponding mock server. The \emph{Workload Generator} then submits scenario prompts sequentially, recording the start and end timestamp and full response for each interaction. Throughout execution, the \emph{Monitoring Module} collects metrics such as GPU utilization, memory usage, and power consumption. These are forwarded to a time-series backend supporting time-range queries during post-processing.

After execution completes, the \emph{Evaluation Engine} compares each trace against the gold-standard solution defined for the corresponding scenario and granularity level. Based on the observed tool calls, argument values, call ordering, and any failures or timeouts, it computes the functional metrics defined in Sec.~\ref{sec:impl}. Using the recorded execution interval, the engine retrieves selected resource metrics 
from the monitoring backend and integrates them with the trace-level results. 
The engine emits an \emph{Evaluation Report} capturing both the agent's functional behavior and its underlying resource footprint, enabling \textsc{MCP-GRANITE} to support reproducible and resource-aware benchmarking across models, domains, and tool granularities.

\subsection{Granularity Modeling and Experiment Specification}
\label{sec:spec}

The central design variable in \textsc{MCP-GRANITE} is the \emph{tool-interface granularity level}, which controls how the same domain operations are partitioned into MCP-exposed tools. {The levels systematically sample the spectrum of granularity recognized in API and service design}~\cite{zimmermann2022patterns}, defined along two dimensions: the \emph{number of tools visible to the agent} and the \emph{amount of operational intent each call must encode}. 
The tools for each level of granularity are created within the framework, as detailed in Sec.~\ref{sec:impl} and Sec.~\ref{sec:setup}, with code available in our repository~\cite{edgetoolbench}. Users can extend the benchmark with additional domains, scenarios, and granularity designs while retaining the same controlled comparison methodology. Each level is implemented as an alternative tool-exposure module over the same underlying domain logic, rather than generated automatically.
Then, they specify their scenarios through a YAML-based \textit{scenario} model, shown in Fig.~\ref{fig:scenario}. Each scenario file includes a natural-language prompt and a per-granularity \emph{gold standard}, which defines the expected sequence of tool calls and arguments required to complete the task. {The same user prompt is used across all levels, isolating the effect of tool-interface granularity from task complexity.}


To illustrate the rationale behind the granularity levels, consider a smart-home scenario (Fig.~\ref{fig:scenario}), where the agent is asked to activate a night security protocol. This requires setting the living room to night mode, locking the doors, enabling camera recording with night vision, checking the front-door sensor, reviewing events, and creating automation rules that trigger an alarm if a door opens after~11pm. 
\textsc{MCP-GRANITE} spans four granularity levels (L4~to~L1) that progressively consolidate the same functionality into broader tools.

\begin{figure}[t]
\centering
\begin{minipage}{\columnwidth}
\begin{lstlisting}[
  basicstyle=\scriptsize\ttfamily,
  frame=single,
  showstringspaces=false,
  breaklines=true,
  columns=fullflexible]
id: smarthome-007
domain: smarthome
difficulty: hard
user_prompt: Activate the full night security protocol. 
Set the living room to night mode, lock all doors, enable 
camera recording with night vision, check front door sensor, 
review recent events, and create a rule that sounds the 
alarm if any door opens after 11pm.
gold_standard:
  L4: # 8 expected calls using Level-4 single-purpose tools
    - tool: set_device
      arguments: { device_id: "DV003", settings: {...} }
    - tool: read_sensor
      arguments: { sensor_id: "SN005" }
    - tool: create_automation_rule
      arguments: { name: "Night door alarm", ... }
      # ... 5 more calls
  L3: # 3 expected calls using Level-3 task-level tools
    - tool: set_room_mode
      arguments: { location: "living room", mode: "night" }
    - tool: get_room_status
      arguments: { location: "front door" }
    - tool: manage_automation
      arguments: { action: "create", ... }
  L2: # 5 expected calls using Level-2 query/control tools
    - tool: smart_home_control
      arguments: { action: "set_mode", ... }
    - tool: smart_home_query
      arguments: { action: "sensor_reading", ... }
      # ... 3 more calls
  L1: # 5 expected calls using the Level-1 monolithic tool
    - tool: smart_home
      arguments: { action: "set_mode", ... }
    - tool: smart_home
      arguments: { action: "sensor_reading", ... }
      # ... 3 more calls
\end{lstlisting}
\end{minipage}
\caption{Smart-home scenario excerpt.}
\label{fig:scenario}
\vspace*{-1.8\baselineskip}
\end{figure}

 {At \textbf{Level~4 (L4)}, the developer designs \emph{one dedicated tool per operation} (e.g., \texttt{set\_device}, \texttt{read\_sensor}, etc.). This granularity level exposes 8-10 single-purpose primitive tools (8 in the smart-home scenario), so the agent must select among many candidates but each call requires only a few, straightforward arguments. This is analogous to fine-grained CRUD (Create, Read, Update, Delete) APIs, where each function performs only one action. 

\textbf{Level~3 (L3)} applies \emph{task-class grouping}, organizing tools into task-level granularity such as \texttt{set\_room\_mode} and \texttt{manage\_automation}. In our benchmark, L3 groups tools into four functional categories: status querying, action execution, configuration, and analysis.
In this level, the agent selects among fewer tools, but each call must specify both the target entity and the desired operation, requiring more complex arguments. This is analogous to resource-oriented or aggregate endpoints, where related operations are grouped under a common interface.

At \textbf{Level~2 (L2)}, the interface reduces to two tools, applying \emph{read/write separation}. For example, in smart-home scenario the interface has \texttt{smart\_home\_query} and \texttt{smart\_home\_control}, separating read from write operations. 
The operation formerly encoded in the tool name is moved into an explicit \texttt{action} parameter, such as \texttt{action: ``set\_mode"}, so most operational intent now resides in the argument schema, increasing argument complexity. This is analogous to Command-Query Responsibility Segregation~(CQRS), a standard pattern in distributed systems that separates data retrieval from state modification.
 
Lastly, \textbf{Level~1 (L1)} introduces a \emph{single-entry-point} design, exposing one monolithic tool through which all operations are accessed. In our scenario, a single \texttt{smart\_home} tool handles everything.
The tool name carries no operational meaning, and the agent must encode the operation type, target entity, and all parameters entirely through arguments. This is analogous to the facade or API gateway pattern, where a single interface dispatches to internal logic based on request parameters.
}

The four levels trace a principled path along the consolidation spectrum. L4 separates by individual operation, L3 by task class, L2 by data-flow direction, and L1 removes all separation. This produces a non-monotonic difficulty profile, as consolidation reduces the tool-selection burden but progressively increases the argument-construction burden. 
Additionally, in our configuration, \textit{gold standards} are defined separately for each granularity level, enabling the benchmark to evaluate correct tool selection and argument accuracy at each level. For instance, the 8 fine-grained calls required at L4 may collapse to 3 calls at L3, while L1 may still require repeated invocations of the same monolithic tool. Thus, consolidation may reduce the number of exposed tools but not the number of calls required, distinguishing interface size from task execution length.

\begin{figure}[t]
\centering
\begin{minipage}{\columnwidth}
\begin{lstlisting}[
  basicstyle=\scriptsize\ttfamily,
  frame=single,
  showstringspaces=false,
  breaklines=true,
  columns=fullflexible]
models:        [llama-3.2, granite-4, gpt-oss, xlam-2, qwen-3, ...]
domains:       [agriculture, robotics, energy, industrial, ...]
granularities: [L4, L3, L2, L1] # Level 4 to Level 1
repetitions:   3

# Agent-level controls
max_agent_turns:       10
agent_timeout_seconds: 360.0
temperature:           0.0
max_tokens:            4096
\end{lstlisting}
\end{minipage}
    \vspace*{-.3\baselineskip}
\caption{Experiment configuration.}
\label{fig:experiment}
    \vspace*{-1.2\baselineskip}
\end{figure}

Lastly, an experiment is specified in a single YAML file~(Fig.~\ref{fig:experiment}) that declares the evaluated models, target domains, granularity levels, and repetitions, together with agent-level controls. The agent-level controls define the maximum number of tool-use steps allowed per task, the execution timeout, the sampling temperature that regulates output randomness, and the maximum generation length for each model response.
Internally, the framework enumerates the full Cartesian product of models $\times$ domains $\times$ granularities $\times$ scenarios and executes each combination for the configured number of repetitions.

\section{Implementation Details}
\label{sec:impl}

\noindent \textbf{MCP Server Implementation: }
The MCP Mock Server is implemented as a FastMCP server~\cite{anthropic2024mcp}, allowing the framework to simulate realistic MCP-based tool interaction while preserving experimental control. The implementation follows a layered design that separates domain representation, state management, and tool exposure. At the lowest layer, each domain defines a set of classes describing the entities manipulated by the server, such as sensors in agriculture, patients in healthcare, or vehicles in fleet management. On top of this layer, each domain provides a set of functions that encapsulates its business logic and operates over an in-memory store maintaining the domain state (e.g., current sensor readings, device configurations, and operational statuses).
Then, a tool-exposure layer is implemented through separate server modules, one per granularity level. Across all four levels, the underlying domain logic and data store remain identical. The only difference lies in the schema and abstraction level of the tools exposed to the agent, ensuring that tool-interface granularity is the sole experimental variable.
To ensure reproducibility, the store is initialized with deterministic test data so that repeated executions of the same scenario observe identical conditions. To prevent cross-run interference, the agent harness launches a fresh MCP server container for each experimental condition, ensuring that no state persists between runs and that every scenario starts from a clean and deterministic environment.

\noindent \textbf{Agentic Service: }
The \emph{Agentic Service} is responsible for instantiating the evaluated LLM and mediating its interaction with the MCP-based tool environment. A key property of this service is that tool discovery is performed dynamically at runtime: at the beginning of each session, the agent retrieves the list of available tools through MCP's \texttt{tools/list} method, rather than relying on statically defined or hard-coded specifications. This closely reflects realistic MCP deployments, in which agents must adapt to the tool schemas exposed by the connected server at execution time rather than operate over a fixed, preconfigured interface.
The service is implemented using Google's Agent Development Kit~(ADK)~\footnote{https://google.github.io/adk-docs/}, with \texttt{McpToolset} handling communication with the MCP server and LiteLLM~\footnote{https://github.com/BerriAI/litellm} providing a uniform abstraction for model inference. This combination allows the same benchmark pipeline to evaluate diverse LLMs through a common OpenAI-compatible interface, without requiring modifications to the orchestration logic. As a result, different model families and sizes can be compared under the same interaction workflow and with identical dynamically discovered tool schemas, ensuring a reproducible execution protocol across all models.

\noindent \textbf{Evaluation \& Utilization Metrics: }
To assess agent performance, each execution is compared against the gold-standard solution defined for the tool-granularity level. Let $G$ denote the gold sequence of tool calls and $\hat{G}$ the sequence produced by the agent. \textit{Tool Selection F1} is computed over the multiset of invoked tool names, ignoring order. Precision is the fraction of predicted tool calls that match tools in $G$, recall is the fraction of gold tool calls recovered in $\hat{G}$, and F1 is their harmonic mean. \textit{Argument Accuracy}~(ArgAcc) measures the fraction of required gold argument key-value pairs correctly produced after aligning predicted and gold calls of the same tool in sequence order. Missing calls, arguments, or mismatched values are counted as incorrect. \textit{Task Completion}~(TC) is a binary metric equal to 1 when the executed trace satisfies the scenario objective according to the gold-standard state and output conditions, and 0 otherwise. Confidence Intervals (CIs) use 2,000 model-level bootstrap resamples after per-model averaging. Comparisons against L4 use paired resampling with Benjamini-Hochberg correction over 9 tests.

We also report robustness and efficiency metrics. The \textit{zero-tool-call rate} is the fraction of runs where the agent returns a final answer without invoking any tool. The \textit{error rate} captures runs terminated by parsing errors, invalid tool calls, runtime exceptions, or schema violations, while the \textit{timeout rate} captures runs exceeding the execution budget. 
Efficiency is measured through wall-clock time, number of tool calls, and energy per successful task, computed for aggregate results as total GPU energy, estimated from mean GPU power and execution time, divided by the number of completed tasks. 
As evaluation uses controlled mock servers, these measurements characterize agent-side execution rather than end-to-end latency or energy of live IoT deployments.
To capture the mean GPU power, we deployed a monitoring stack using Prometheus and NetData on edge node\footnote{https://prometheus.io/ \& https://www.netdata.cloud/}, supplemented by a custom \texttt{nvidia-smi} wrapper sampling at 5s intervals.

{Beyond the metrics reported here, the framework collects a broader set of trace-level and resource metrics including token counts, redundant-call rates, and node-level resource telemetry, such as GPU utilization, memory, power draw, temperature, and many more. 
Due to space constraints, we focus on GPU utilization, power consumption, and wall-clock time as the metrics most relevant to the granularity analysis (Sec.}~\ref{sec:resourcesenergy}).

%
\vspace*{-.1\baselineskip}
\section{Experimental Evaluation}
\label{sec:setup}



We instantiate \textsc{MCP-GRANITE} across 9 IoT domains organized into three application areas~\cite{lin2025edgellmsurvey}: industrial automation (\textit{industrial}, \textit{robotics}, \textit{warehouse}), smart environments (\textit{smart home}, \textit{surveillance}, \textit{energy}), and field operations (\textit{agriculture}, \textit{fleet}, \textit{healthcare}). Each domain is operationalized by an in-memory data store that holds the entities the agent interacts with, such as sensors, devices, patients, or vehicles, each identified by a fixed ID (e.g., \texttt{SN005}, \texttt{DV003}). For every scenario, a gold standard specifies the expected sequence of tool calls and their arguments at each granularity level. These gold standards were verified by running them against the store and confirming that the correct final state was reached. Scenarios are organized into three difficulty tiers based on how many tool calls and distinct entities they require: \emph{easy} (1-2 calls, single entity), \emph{medium} (4-5 calls, 2-3 entities), and \emph{hard} (9-10 calls, 4+ entities). Each of the 9 domains contributes 3 scenarios per tier, yielding 81 scenarios in a balanced design. The complete scenario definitions and benchmark construction details are publicly available in the paper's repository~\cite{edgetoolbench}.

We evaluate nine models spanning 268M to 20.9B parameters (Table~\ref{tab:overview}), served through Ollama\footnote{https://ollama.com/}. The selection includes both general-purpose and function-calling variants. Quantization varies by model: FunctionGemma uses 8-bit, xLAM-2 uses 3-bit, Qwen3 uses 6-bit, GPT-OSS uses MXFP4 (block-scaled 4-bit), and the remaining five use 4-bit. These affect per-model performance, as higher-bit quantization preserves more weight fidelity but increases memory use and inference time. All experiments ran on a GPU-enabled edge workstation with an NVIDIA Tesla T4, 16GB VRAM, 128GB RAM, and a multi-core CPU, representing an edge-server or gateway-class deployment~\cite{lin2025edgellmsurvey}. 
Our full-factorial design combines 9 models, 9 domains, 4 granularity levels, 9 scenarios, and 3 repetitions, yielding 8,748 runs (972 per model). 
Lastly, each configuration had the same agent-level controls as Fig.~\ref{fig:experiment}.

\subsection{Overall Model Performance}

Table~\ref{tab:overview} summarizes mean performance across all scenarios. Qwen3 (14.8B) achieves the best performance, with 0.762 task completion, 0.857 F1, and 0.531 argument accuracy, but at a latency of 164.5s, driven by its large parameter count and its built-in reasoning capabilities, which produce extended thinking traces before each invocation. Llama3.2 (3.2B) is the second-best model, reaching 0.579 task completion and 0.760 F1 in only 6.7s with zero errors, making it 24.5$\times$ faster than Qwen3 while retaining about 76\% of its task completion. Mistral-Nemo also performs competitively, with 0.488 task completion, 0.676 F1, and a low error rate of 1.9\%. In contrast, GPT-OSS and xLAM-2 show very high error rates of 55.6\% and 52.8\%, despite their larger size. 
GPT-OSS is a Mixture-of-Experts model with 20.9B total parameters but only $\approx$3.6B active per forward pass. Its errors are dominated by malformed tool-call outputs and schema violations. xLAM-2, which uses aggressive 3-bit quantization, shows similar structured-output failures, particularly when tool schemas are discovered at runtime rather than provided statically.
At the other end, FunctionGemma is the fastest model at 2.9s, but its performance remains far below the stronger models.

\begin{table}[t]
\centering
\caption{Models' overall performance. TC with 95\% CI. F1, ArgAcc as mean${\pm}$std. G/G+T/G+C/FC = General / +thinking / +chat / Function calling. Calls = mean tool calls per scenario.}
\label{tab:overview}
\scriptsize
\setlength{\tabcolsep}{2pt}
\renewcommand{\arraystretch}{0.85}
\begin{tabular}{@{}lrcccrcc@{}}
\toprule
\textbf{Model} & \textbf{Params} & \textbf{TC [95\% CI]} & \textbf{F1} & \textbf{ArgAcc} & \textbf{Time(s)} & \textbf{Err\%} & \textbf{Calls}\\
\midrule
Qwen3\,\textsuperscript{G+T}        & 14.8B & $\mathbf{.76\,[.74,.79]}$ & $\mathbf{.86{\pm}.26}$ & $\mathbf{.53{\pm}.36}$ & 164.5 & 5.5  & 3.7\\
Llama3.2\,\textsuperscript{G}       & 3.2B  & $.58\,[.55,.61]$ & $.76{\pm}.33$ & $.29{\pm}.35$ & 6.7   & 0.0  & 2.0\\
Mistral-Nemo\,\textsuperscript{G}   & 12.2B & $.49\,[.46,.52]$ & $.68{\pm}.39$ & $.34{\pm}.35$ & 17.7  & 1.9  & 1.9\\
Ministral-3\,\textsuperscript{G}    & 13.9B & $.42\,[.39,.45]$ & $.50{\pm}.45$ & $.26{\pm}.33$ & 27.4  & 10.5 & 2.7\\
Granite4\,\textsuperscript{G}       & 3.4B  & $.40\,[.37,.43]$ & $.63{\pm}.38$ & $.27{\pm}.34$ & 7.4   & 0.2  & 1.4\\
Hermes3\,\textsuperscript{G+C}      & 8.0B  & $.39\,[.36,.42]$ & $.75{\pm}.35$ & $.31{\pm}.34$ & 8.4   & 0.0  & 2.2\\
GPT-OSS\,\textsuperscript{G}        & 20.9B & $.34\,[.31,.37]$ & $.66{\pm}.31$ & $.36{\pm}.37$ & 25.4  & 55.6 & 3.7\\
xLAM-2\,\textsuperscript{FC}        & 8.0B  & $.25\,[.22,.28]$ & $.63{\pm}.36$ & $.35{\pm}.35$ & 9.2   & 52.8 & 2.9\\
FuncGemma\,\textsuperscript{FC}     & 268M  & $.19\,[.16,.21]$ & $.37{\pm}.42$ & $.10{\pm}.22$ & 2.9   & 0.5  & 1.0\\
\bottomrule
\end{tabular}
\vspace*{-1.8\baselineskip}
\end{table}

\smallskip\noindent
\textit{\textbf{Key takeaway:} Model size is not a reliable predictor of tool-use performance. Qwen3 achieves the highest accuracy, but Llama3.2 offers a better efficiency–robustness trade-off, while GPT-OSS and xLAM-2 show that larger or function-calling models do not generalize to dynamic tool interfaces.}

\begin{figure}[h]
  \vspace*{-.6\baselineskip}
  \centering
  \begin{minipage}[t]{0.5\textwidth}
    \centering
    \includegraphics[width=0.75\linewidth]{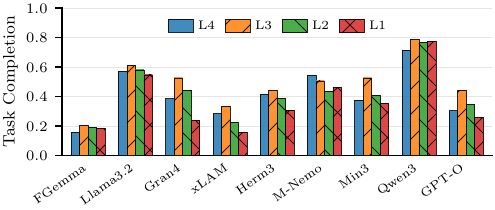}
    \caption{Task completion by model and levels L1-L4}
    \label{fig:granularity_tc}
  \end{minipage}\hfill
    
    \vspace*{-0.5\baselineskip}
\end{figure}

\subsection{Effect of Tool-Interface Granularity}
\label{sec:granularity}

Across nearly all models, L3, representing moderate consolidation, is the most effective granularity configuration~(Fig.~\ref{fig:granularity_tc}), achieving the highest task completion for 8 of 9 models. Mistral-Nemo is the only exception, where L4 performs slightly better. 
L3 reaches a task completion of 0.49, a $+16.4\%$ improvement over L4 and $+33.6\%$ over L1, computed from unrounded means  (Table~\ref{tab:granularity_stats}). 
Its zero-tool-call rate is slightly higher than L4 (10.7\% vs.\ 10.2\%), indicating that L3 does not improve every individual metric. 
It also yields the highest argument accuracy (0.40) and an F1 of 0.66, while reducing execution time.
Paired bootstrap tests confirm that the L3 improvements are significant (Table~\ref{tab:granularity_stats}). 
To verify this, we fit a logistic regression including model size, quantization, domain, and difficulty, which confirms that L3 independently improves completion odds by 1.36$\times$ (CI [1.12, 1.66], $p{=}0.002$), while L1 reduces them by 22\% ($p{=}0.009$).

In contrast, the fully consolidated L1 setting substantially degrades performance, with task completion dropping to 0.36 ($-12.9\%$ vs L4) and zero-tool-call rates rising from 10.2\% at L4 to 28.2\% (Table~\ref{tab:granularity_stats}). Collapsing the interface into a single monolithic tool forces the model to choose among many action types and populate a larger schema in every call. The effect is most pronounced for weaker models (FunctionGemma and xLAM-2), but also reduces task completion for Granite4 and Hermes3, confirming that excessive consolidation simplifies the interface at the cost of execution reliability.


\begin{table}[t]
\centering
\caption{TC, F1, ArgAcc as mean [95\% CI] from 2,000 model-level bootstrap resamples. Bold = best. $\dagger$: $p<0.05$; $\ddagger$: $p<0.01$ vs.\ L4 (BH-adjusted paired bootstrap).}
\label{tab:granularity_stats}
\scriptsize
\setlength{\tabcolsep}{2pt}
\renewcommand{\arraystretch}{0.95}
\begin{tabular}{@{}lcccc@{}}
\toprule
\textbf{Granularity} & \textbf{TC [95\% CI]} & \textbf{F1 [95\% CI]} & \textbf{ArgAcc [95\% CI]} & \textbf{Zero\% / Time(s)} \\
\midrule
L4 (8-10)            & 0.42\,[.31,.52]                       & 0.59\,[.51,.67]                       & 0.20\,[.15,.27]                       & 10.2\% / 33.0 \\
L3 (4)$^\ddagger$     & \textbf{0.49}\,[.39,.59]              & \textbf{0.66}\,[.58,.74]$^\ddagger$   & \textbf{0.40}\,[.33,.48]$^\ddagger$   & 10.7\% / 31.0 \\
L2 (2)                & 0.42\,[.32,.53]                       & 0.71\,[.58,.83]$^\ddagger$            & 0.38\,[.28,.45]$^\ddagger$            & 14.6\% / 30.4 \\
L1 (1)$^\dagger$      & 0.36\,[.25,.50]                       & 0.63\,[.52,.73]                       & 0.26\,[.19,.34]$^\ddagger$            & \textbf{28.2\%} / 25.1 \\
\bottomrule
\end{tabular}
\vspace*{-1.0\baselineskip}
\end{table}

\begin{table}[t]
\centering
\caption{Granularity effects by model size class.}
\label{tab:size_gran}
\scriptsize
\begin{tabular}{@{}llcccc@{}}
\toprule
\textbf{Metric} & \textbf{$<$8B $|$ $\geq$8B} & \textbf{L4} & \textbf{L3} & \textbf{L2} & \textbf{L1} \\
\midrule
TC     & $<$8B $|$ $\geq$8B & .395 $|$ .440 & \textbf{.465 $|$ .506} & .422 $|$ .429 & .347 $|$ .385 \\
F1     & $<$8B $|$ $\geq$8B & .540 $|$ .619 & .646 $|$ .673 & \textbf{.675 $|$ .739} & .548 $|$ .677 \\
ArgAcc & $<$8B $|$ $\geq$8B & .121 $|$ .245 & \textbf{.313 $|$ .451} & .299 $|$ .420 & .168 $|$ .316 \\
\bottomrule
\end{tabular}
\vspace*{-2.0\baselineskip}
\end{table}

Argument accuracy benefits most consistently from consolidation. Moving from L4 to L3 nearly doubles mean accuracy, from 0.20 to 0.40 (a $+99.5\%$ improvement). L2 remains similarly high at 0.38 ($+85.8\%$ over L4). 
Moderate consolidation reduces tool-identification burden and helps models construct more accurate arguments. Task completion also improves for 8 of 9 models, showing that the L3 advantage generalizes.

A more nuanced pattern appears in tool-selection F1. Although task completion peaks at L3, the highest mean F1 is achieved by L2 (0.71, $+20.8\%$ over L4). This likely reflects the smaller decision space at that granularity, where choosing between two broad tools is easier than selecting among four or more specialized ones. 
By contrast, the F1 advantage of L1 over L4 is small (+6.7\%) and is not significant under the paired bootstrap comparison (Table~\ref{tab:granularity_stats}), so collapsing the interface to a single tool does not preserve the F1 gains of moderate consolidation.
The higher F1 of L2 does not translate into the highest task completion, since successful execution depends on accurately specifying arguments within the selected tool. This suggests that correct tool selection alone is insufficient and that argument construction remains a bottleneck. This is consistent with function-calling evaluations that distinguish selection correctness from argument accuracy and find that they can fail independently~\cite{patil2024bfcl, song2025mcpgauge}.

\smallskip
\noindent \textit{\textbf{Key takeaway}: Moderate tool consolidation with L3 is the most effective design choice, delivering the best balance of task completion, argument accuracy, and latency. Over-consolidation increases failures and confirms that argument construction, not tool selection, is the main bottleneck.}

\subsection{Scaling Analysis}
\label{sec:scaling}

\begin{figure}[h]
  \centering
  \begin{minipage}[t]{0.5\textwidth}
    \centering
    \includegraphics[width=0.85\linewidth]{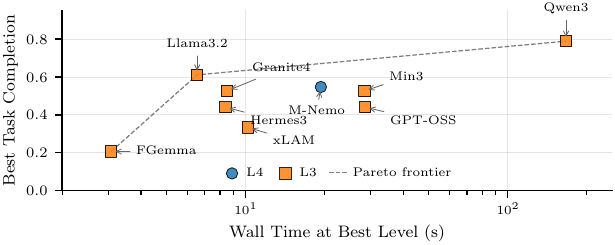}
    \caption{Best task completion vs. time (log~scale)}
    \label{fig:pareto}
  \end{minipage}\hfill
  \begin{minipage}[t]{0.5\textwidth}
    \centering
    \includegraphics[width=0.85\linewidth]{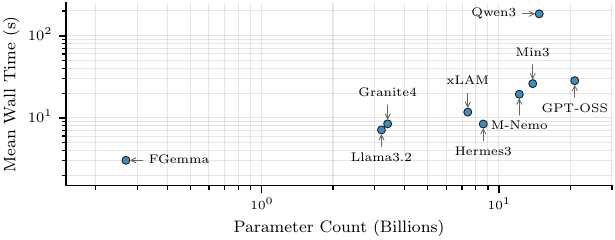}
    \caption{Mean wall time vs.\ model size (log-log).}
    \label{fig:walltime}
  \end{minipage}
  \vspace*{-.5\baselineskip}
\end{figure}

Fig.~\ref{fig:pareto} plots each model at the granularity level where it achieves its highest task completion against the corresponding wall time. Three models define the Pareto frontier: FunctionGemma (3.1\,s, TC\,=\,0.21), Llama\,3.2 (6.6\,s, TC\,=\,0.61), and Qwen\,3 (167\,s, TC\,=\,0.79), with all others dominated on at least one axis. 
Among the dominated models, Hermes\,3 and Granite\,4 are closest to the frontier, offering competitive task completion at moderate latency. In contrast, GPT-OSS and Ministral\,3 incur much higher wall times without proportional gains, reinforcing that model scale alone does not determine the best edge-deployment trade-off.

\vspace*{-.5\baselineskip}

\begin{figure*}[t]
  \centering
  \begin{minipage}[t]{0.5\textwidth}
    \centering
    \includegraphics[width=0.75\linewidth]{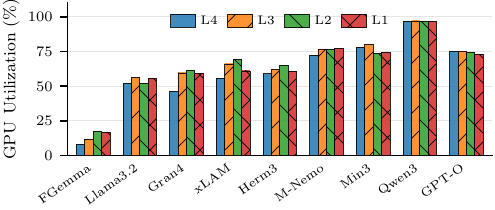}
    \caption{Mean GPU utilization by model \&
    level L4-L1.}
    \label{fig:resource_bars}
  \end{minipage}\hfill
  \begin{minipage}[t]{0.5\textwidth}
    \centering
    \includegraphics[width=0.75\linewidth]{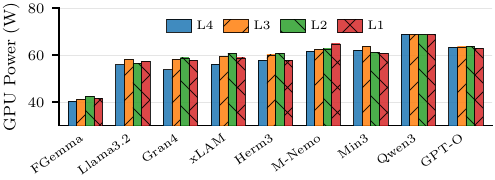}
    \caption{Mean GPU power draw by model \& level L4-L1. }
    \label{fig:gpu_power}
  \end{minipage}
  \vspace*{-1\baselineskip}
\end{figure*}

Within the nine evaluated models, parameter count is more strongly associated with argument accuracy and latency than with task completion. Parameter count correlates positively with argument accuracy ($\rho{=}0.686$, $p{=}0.041$), indicating higher argument accuracy among larger models in this sample, and even more strongly with wall-clock time ($\rho{=}0.946$, $p{<}0.001$). For example, average execution time increases from 2.9s for FunctionGemma to 164.5s for Qwen3, a 56$\times$ increase. Fig.~\ref{fig:walltime} further shows that latency grows roughly linearly with parameter count on the log-log scale, except for Qwen3, which is slower due to extended reasoning traces, and GPT-OSS, which is faster than its 20.9B parameters suggest because its MoE architecture activates only $\approx$3.6B parameters per forward pass. As a sensitivity check, excluding Qwen3 and GPT-OSS reduces the argument-accuracy association to $\rho{=}0.360$ ($p{=}0.427$), while the latency association remains strong ($\rho{=}0.991$, $p{<}0.001$). 
In contrast, parameter count is only weakly and non-significantly associated with task completion ($\rho{=}0.285$, $p{=}0.458$), showing that scale alone does not determine end-to-end agent success.

To examine if tool-interface granularity depends on model scale, Table~\ref{tab:size_gran} groups models into small ($<8$B) and large ($\geq8$B). Both benefit from moderate consolidation, with L3 achieving the best task completion. The gain is slightly higher for small models, 17.7\% versus 15.0\% for large models. At L2, small models improve by 6.8\%, while large models decrease by 2.5\%, suggesting that smaller models benefit more from a reduced tool-selection space, whereas larger models better exploit the structure preserved at L3. L1 degrades both groups, confirming that over-consolidation is harmful across scale. These trends may also reflect quantization and training differences, since FunctionGemma uses 8-bit quantization while xLAM-2 uses aggressive 3-bit quantization, which may reduce structured-output reliability. 
Thus, these size associations are descriptive and should not be interpreted independently of quantization, architecture, and tool-use alignment.

\smallskip\noindent \textit{\textbf{Key takeaway}: 
Model size alone is a poor predictor of end-to-end agent success. Its clearest association is with latency, while the observed relationship with argument accuracy is less robust. Smaller models appear more sensitive to tool-interface design and benefit more from moderate consolidation.
}

\vspace*{-0.2\baselineskip}
\begin{table}[h]
\centering
\caption{Joules per task under L3. Power, Time, and J/task are $\mathrm{mean}{\pm}\mathrm{std}$. TC is shown with its 95\% confidence interval.}
\label{tab:energy}
\scriptsize
\setlength{\tabcolsep}{3pt}
\begin{tabular}{@{}lcccc@{}}
\toprule
\textbf{Model} & \textbf{Power (W)} & \textbf{Time (s)} & \textbf{TC [95\% CI]} & \textbf{J/task} \\
\midrule
Llama3.2   & $58.2{\pm}14.5$ & $6.5{\pm}2.9$    & $0.61\,[.56,.66]$ & $621{\pm}585$ \\
Granite4   & $58.1{\pm}13.5$ & $8.5{\pm}7.3$    & $0.53\,[.46,.59]$ & $937{\pm}1{,}219$ \\
Hermes3    & $60.1{\pm}13.2$ & $8.4{\pm}4.2$    & $0.44\,[.38,.50]$ & $1{,}148{\pm}1{,}439$ \\
FuncGemma    & $41.1{\pm}10.0$ & $3.1{\pm}0.7$    & $0.21\,[.16,.26]$ & $618{\pm}1{,}232$ \\
xLAM-2       & $59.6{\pm}12.9$ & $10.2{\pm}9.3$   & $0.33\,[.28,.39]$ & $1{,}826{\pm}3{,}103$ \\
Mistral-Nemo & $62.4{\pm}11.0$ & $17.6{\pm}18.5$  & $0.51\,[.44,.57]$ & $2{,}170{\pm}3{,}157$ \\
Ministral-3  & $63.6{\pm}10.8$ & $28.4{\pm}33.5$  & $0.53\,[.46,.59]$ & $3{,}427{\pm}5{,}220$ \\
GPT-OSS      & $63.5{\pm}9.2$  & $28.6{\pm}20.4$  & $0.44\,[.38,.50]$ & $4{,}128{\pm}5{,}547$ \\
Qwen3      & $68.8{\pm}1.6$  & $167.0{\pm}94.9$ & $0.79\,[.74,.84]$ & $14{,}544{\pm}11{,}173$ \\
\bottomrule
\end{tabular}
  \vspace*{-1.0\baselineskip}
\end{table}

\begin{table}[h]
\centering
\caption{Efficiency (TC/second, mean $\pm$ std) by model and granularity. Bold = best granularity per model.}
\label{tab:efficiency}
\scriptsize
\setlength{\tabcolsep}{2pt}
\begin{tabular}{@{}lcccc@{}}
\toprule
\textbf{Model} & \textbf{L4} & \textbf{L3} & \textbf{L2} & \textbf{L1} \\
\midrule
Llama3.2    & $.081{\pm}.101$ & $.093{\pm}.099$ & $.078{\pm}.094$ & $\mathbf{.100{\pm}.126}$ \\
FuncGemma     & $.053{\pm}.129$ & $.067{\pm}.134$ & $\mathbf{.070{\pm}.152}$ & $.069{\pm}.156$ \\
Granite4    & $.046{\pm}.083$ & $\mathbf{.062{\pm}.085}$ & $.060{\pm}.089$ & $.046{\pm}.074$ \\
Hermes3     & $.049{\pm}.081$ & $\mathbf{.053{\pm}.077}$ & $.043{\pm}.071$ & $.040{\pm}.069$ \\
xLAM-2        & $.024{\pm}.062$ & $\mathbf{.033{\pm}.066}$ & $.026{\pm}.054$ & $.024{\pm}.048$ \\
Mistral-Nemo  & $.028{\pm}.054$ & $\mathbf{.029{\pm}.054}$ & $.026{\pm}.048$ & $.027{\pm}.052$ \\
Ministral-3   & $.014{\pm}.048$ & $\mathbf{.019{\pm}.063}$ & $.013{\pm}.035$ & $.015{\pm}.056$ \\
GPT-OSS       & $.011{\pm}.032$ & $\mathbf{.015{\pm}.041}$ & $.014{\pm}.035$ & $.013{\pm}.032$ \\
Qwen3       & $.004{\pm}.005$ & $\mathbf{.005{\pm}.007}$ & $.005{\pm}.006$ & $.005{\pm}.007$ \\
\bottomrule
\end{tabular}
\vspace*{-1.5\baselineskip}
\end{table}

\subsection{Resource Utilization and Energy Efficiency}
\label{sec:resourcesenergy}

Fig.~\ref{fig:resource_bars} and Fig.~\ref{fig:gpu_power} summarize GPU utilization and power consumption across models and granularity levels. Both metrics are primarily driven by model size: GPU utilization ranges from 8-17\% for FunctionGemma to 96-97\% for Qwen3, while GPU power increases more modestly, from roughly 40W to 69W. Power scales sub-linearly with size, as Qwen3 is about 55$\times$ larger than FunctionGemma but draws only 1.7$\times$ more power, because smaller models leave GPU compute capacity idle. 
VRAM usage is dominated by the loaded model weights and therefore varies little across granularity levels.

Granularity affects utilization mainly when the GPU is under-saturated. For large models, utilization changes by less than 3\% across levels, while smaller models show larger swings (e.g., FunctionGemma rising from 8.3\% at L4 to 17.6\% at L2). GPU power changes by at most $\sim$5W per model. Hardware cost is thus dominated by model choice, with granularity modulating under-saturated edge models.

To assess end-to-end efficiency, we compute GPU energy per successful task from mean GPU power and wall-clock time (Table~\ref{tab:energy}). Llama3.2 is the most favorable operating point at 621J/task under L3, whereas Qwen3 requires 14,544J/task, a 23$\times$ higher cost. FunctionGemma appears competitive (618J/task), but completes only 20.6\% of tasks, so much of its energy is wasted on failed tasks, while Llama3.2 consumes similar energy but completes 3$\times$ more tasks.
Larger models improve capability at disproportionate time and energy cost, while mid-sized models offer the best trade-off~(Table~\ref{tab:efficiency}).

\smallskip
\noindent
\textit{\textbf{Key takeaway}: Resource and energy efficiency are driven by model choice, with granularity as a secondary factor. Longer execution times dominate energy cost since power scales sub-linearly with size. Granularity mainly affects under-saturated small and mid-sized models.}

\vspace*{-.3\baselineskip}
\subsection{Domain Difficulty and Failure Analysis}
\label{sec:domain_failures}
\vspace*{-.2\baselineskip}

\begin{figure}[t]
  \centering
  \includegraphics[width=0.85\columnwidth]{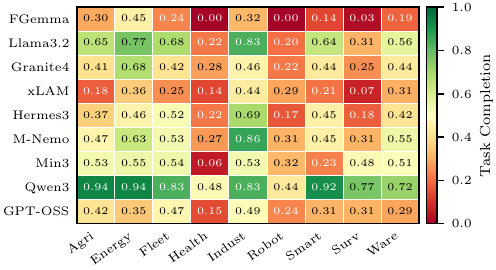}
  \caption{Task completion by model and domain}
  \label{fig:domain_heatmap}
    \vspace*{-1.4\baselineskip}
\end{figure}

Fig.~\ref{fig:domain_heatmap} shows strong variation across both models and domains: Qwen3 performs best across most domains, while healthcare (0.202) and robotics (0.244) have the lowest mean completion and industrial (0.608) and energy (0.596) the highest.
Despite these differences, tool-selection F1 remains stable (0.614-0.667), indicating that models generally identify the right tools but fail during multi-step reasoning and argument construction. This pattern holds at the scenario level, where task completion ranges from 0.852 (warehouse-001) to 0.009 (healthcare-003), yet even the hardest scenarios achieve moderate F1 (0.596-0.671). Across all 8,748 conditions, 13.6\% result in errors and 1.3\% in timeouts. Failures are highly model-dependent: GPT-OSS and xLAM-2 produce error rates of 55.6\% and 52.8\%, while Llama3.2 and Hermes3 produce none. Tool-interface granularity also affects robustness, with L3 yielding the lowest error rate (11.4\%), followed by L4 (12.9\%), L2 (14.1\%), and L1 (15.8\%). The degradation at L1 is mainly driven by zero-tool-call behavior (28.2\% of L1 conditions), where models either hallucinate completion without invoking any tool or fall back to memorized tool names that do not match the MCP schema.

\smallskip
\noindent
\textit{\textbf{Key takeaway}: 
Agent performance is limited less by tool identification than by reliable action sequencing. L3 provides the most robust behavior, while L1 produces qualitative breakdown rather than gradual degradation.}

\section{Conclusion and Future Work}
\label{sec:conclusion}

\textsc{MCP-GRANITE} is an extensible, open-source framework for studying MCP tool-interface granularity, evaluated across 9 models and 9 edge/IoT domains on a T4-based platform with controlled mock-server responses. Moderate consolidation (L3, 4 task-class tools) performs best, improving task completion by $16.4\%$ over fine-grained tools and $33.6\%$ over a monolithic interface, primarily through higher argument accuracy. Smaller models benefit most, while hardware cost is driven mainly by model choice: at L3, a 3.2B model uses 23$\times$ less energy per successful task than the best-performing 14.8B model. Over-consolidation reduces task completion by $12.9\%$, with 28.2\% of L1 runs invoking no tool. Overall, our experimental results favor compact, task-semantic interfaces and evaluating granularity alternatives before deployment.

{Our \textbf{future work} will extend the benchmark to live MCP servers, broader model and edge-hardware coverage, and use the collected metrics to model how interface complexity affects agent behavior and formalize tool-granularity design.}

\vspace{.5\baselineskip}
\footnotesize{\noindent\textbf{Acknowledgments:} This work was co-supported by Pharos-CY, funded by the EuroHPC JU (GA: 101263007), with support from the EU's Horizon Europe programme and the Government of the Republic of Cyprus, and by the EU Commission through the AI-DAPT project (HORIZON-CL4-2023-HUMAN-01-01, GA: 101135826). Language was refined using ChatGPT; all content and ideas are the authors’ own.}

\normalsize

\balance
\bibliographystyle{IEEEtran}
\bibliography{references}

@misc{anthropic2024mcp,
  author       = {{Anthropic}},
  title        = {Model Context Protocol: Specification},
  year         = {2024},
  howpublished = {\url{https://modelcontextprotocol.io/specification}},
}

@inproceedings{rabinovich2025robustness,
  title={On the robustness of agentic function calling},
  author={Rabinovich, Ella and Tavor, Ateret Anaby},
  booktitle={TrustNLP},
  year={2025}
}

@article{guo2026learning,
  title={Learning to Rewrite Tool Descriptions for Reliable LLM-Agent Tool Use},
  author={Guo, Ruocheng and Dong, Kaiwen and Gao, Xiang and Das, Kamalika},
  journal={arXiv:2602.20426},
  year={2026}
}

@inproceedings{faghih2025toolpreferences,
  title={Tool Preferences in Agentic LLMs are Unreliable},
  author={Faghih, Kazem and Wang, Wenxiao and Cheng, Yize and Bharti, Siddhant and Sriramanan, Gaurang and Balasubramanian, Sriram and Hosseini, Parsa and Feizi, Soheil},
  booktitle={EMNLP},
  year={2025}
}

@misc{edgetoolbench,
  author = {Paschalides et al.},
  title  = {{MCP-GRANITE Benchmark}},
  year   = {2026},
  url    = {https://github.com/dpasch01/mcp-granite}
}

@inproceedings{wang2025mtu,
  title={Mtu-bench: A multi-granularity tool-use benchmark for large language models},
  author={Wang, Pei and Wu, Yanan and Wang, Zekun and Liu, Jiaheng and Song, Xiaoshuai and Peng, ZY and Zhang, Chenchen and Peng, Junran and Zhang, Ge and Guo, Hangyu and others},
  booktitle={ICLR},
  year={2025}
}

@inproceedings{wu2024toolplanner,
    title = "{T}ool{P}lanner: A Tool Augmented {LLM} for Multi Granularity Instructions with Path Planning and Feedback",
    author = "Wu, Qinzhuo  and
      Liu, Wei  and
      Luan, Jian  and
      Wang, Bin",
    booktitle = "EMNLP",
    year = "2024"
}

@inproceedings{schick2024toolformer,
  author    = {Schick, Timo and Dwivedi-Yu, Jane and Dess{\`i}, Roberto and Raileanu, Roberta and Lomeli, Maria and Hambro, Eric and Zettlemoyer, Luke and Cancedda, Nicola and Scialom, Thomas},
  title     = {Toolformer: Language Models Can Teach Themselves to Use Tools},
  booktitle = {NeurIPS},
  year      = {2024}
}

@inproceedings{yao2023react,
  author    = {Yao, Shunyu and Zhao, Jeffrey and Yu, Dian and Du, Nan and Shafran, Izhak and Narasimhan, Karthik and Cao, Yuan},
  title     = {{ReAct}: Synergizing Reasoning and Acting in Language Models},
  booktitle = {ICLR},
  year      = {2023}
}

@inproceedings{qin2024tool,
  author    = {Qin, Yujia and Liang, Shihao and Ye, Yining and Zhu, Kunlun and Yan, Lan and Lu, Yaxi and Lin, Yankai and Cong, Xin and Tang, Xiangru and Qian, Bill and others},
  title     = {{ToolLLM}: Facilitating Large Language Models to Master 16000+ Real-World {APIs}},
  booktitle = {ICLR},
  year      = {2024}
}

@inproceedings{patil2023gorilla,
  author    = {Patil, Shishir G. and Zhang, Tianjun and Wang, Xin and Gonzalez, Joseph E.},
  title     = {Gorilla: Large Language Model Connected with Massive {APIs}},
  booktitle = {NeurIPS},
  year      = {2023}
}

@inproceedings{patil2024bfcl,
title={The Berkeley Function Calling Leaderboard (BFCL): From Tool Use to Agentic Evaluation of Large Language Models}, 
author={Patil, Shishir G. and Mao, Huanzhi and Cheng-Jie Ji, Charlie and Yan, Fanjia and Suresh, Vishnu and Stoica, Ion and E. Gonzalez, Joseph},
booktitle={ICML},
year={2025},
}

@inproceedings{liu2024agentbench,
  author    = {Liu, Xiao and Yu, Hao and Zhang, Hanchen and Xu, Yifan and Lei, Xuanyu and Lai, Hanyu and Gu, Yu and Ding, Hangliang and Men, Kaiwen and Yang, Kejuan and others},
  title     = {{AgentBench}: Evaluating {LLMs} as Agents},
  booktitle = {ICLR},
  year      = {2024}
}

@inproceedings{zhou2024webarena,
  author    = {Zhou, Shuyan and Xu, Frank F. and Zhu, Hao and Zhou, Xuhui and Lo, Robert and Sridhar, Abishek and Cheng, Xianyi and Ou, Tianyue and Bisk, Yonatan and Fried, Daniel and others},
  title     = {{WebArena}: A Realistic Web Environment for Building Autonomous Agents},
  booktitle = {ICLR},
  year      = {2024}
}

@article{jimenez2024swebench,
  author  = {Jimenez, Carlos E. and Yang, John and Wettig, Alexander and Yao, Shunyu and Pei, Kexin and Press, Ofir and Narasimhan, Karthik},
  title   = {{SWE-bench}: Can Language Models Resolve Real-World {GitHub} Issues?},
  journal = {ICLR},
  year    = {2024}
}

@article{xi2023rise,
  title={The rise and potential of large language model based agents: A survey},
  author={Xi, Zhiheng and Chen, Wenxiang and Guo, Xin and He, Wei and Ding, Yiwen and Hong, Boyang and Zhang, Ming and Wang, Junzhe and Jin, Senjie and Zhou, Enyu and others},
  journal={Sci. China Inf. Sci.},
  year={2025},
  publisher={Springer}
}

@book{zimmermann2022patterns,
  author    = {Zimmermann, Olaf and Stocker, Mirko and L{\"u}bke, Daniel and Zdun, Uwe and Pautasso, Cesare},
  title     = {Patterns for {API} Design: Simplifying Integration with Loosely Coupled Message Exchanges},
  publisher = {Addison-Wesley Professional},
  year      = {2022}
}

@article{myers2016improving,
  author  = {Myers, Brad A. and Stylos, Jeffrey},
  title   = {Improving {API} Usability},
  journal = {Commun. ACM},
  year    = {2016}
}

@inproceedings{zhu2024toolsandbox,
  title={Toolsandbox: A stateful, conversational, interactive evaluation benchmark for llm tool use capabilities},
  author={Lu, Jiarui and Holleis, Thomas and Zhang, Yizhe and Aumayer, Bernhard and Nan, Feng and Bai, Haoping and Ma, Shuang and Ma, Shen and Li, Mengyu and Yin, Guoli and others},
  booktitle={NAACL},
  year={2025}
}

@article{mialon2023augmented,
  title={Augmented language models: a survey},
  author={Mialon, Gr{\'e}goire and Dess{\`\i}, Roberto and Lomeli, Maria and Nalmpantis, Christoforos and Pasunuru, Ram and Raileanu, Roberta and Rozi{\`e}re, Baptiste and Schick, Timo and Dwivedi-Yu, Jane and Celikyilmaz, Asli and others},
  journal={arXiv preprint arXiv:2302.07842},
  year={2023}
}

@inproceedings{yang2024sweagent,
  author    = {Yang, John and Jimenez, Carlos E. and Wettig, Alexander and Narasimhan, Karthik and Yao, Shunyu and Press, Ofir},
  title     = {{SWE-agent}: Agent-Computer Interfaces Enable Automated Software Engineering},
  booktitle = {NeurIPS},
  year      = {2024}
}

@article{song2025mcpgauge,
  author  = {Song, Wentao and Zhong, Hao and Ding, Zhuoer and Xue, Jing and Li, Yiming},
  title   = {Help or Hurdle? Rethinking Model Context Protocol-Augmented Large Language Models},
  journal = {arXiv preprint arXiv:2508.12566},
  year    = {2025}
}

@article{yang2025mcpuniverse,
  author  = {Yang, Le and others},
  title   = {{MCP-Universe}: Benchmarking Large Language Models with Real-World Model Context Protocol Servers},
  journal = {arXiv preprint arXiv:2508.14704},
  year    = {2025}
}

@INPROCEEDINGS{Paramanayakam2025,
  author={Paramanayakam, Varatheepan and Karatzas, Andreas and Anagnostopoulos, Iraklis and Stamoulis, Dimitrios},
  booktitle={DATE}, 
  title={Less is More: Optimizing Function Calling for LLM Execution on Edge Devices}, 
  year={2025},
  volume={},
  number={},}

@misc{mastouri2026restmcpempiricalstudy,
      title={From REST to MCP: An Empirical Study of API Wrapping and Automated Server Generation for LLM Agents}, 
      author={Meriem Mastouri and Emna Ksontini and Amine Barrak and Wael Kessentini},
      year={2026},
      url={https://arxiv.org/abs/2507.16044}, 
}

@article{lin2025edgellmsurvey,
  author  = {Lin, Zhongrui and others},
  title   = {A Review on Edge Large Language Models: Design, Execution, and Applications},
  journal = {ACM Comp. Surveys},
  year    = {2025}
}

\end{document}